\documentclass[prl,aps,twocolumn,showpacs,10pt,superscriptaddress]{revtex4-2}

\usepackage{graphicx}  
\graphicspath{{./Figures/}}
\usepackage{dcolumn}  

\usepackage[nooneline,hang]{subfigure}
\usepackage{amssymb, amsmath}
\usepackage{multibib}
\newcommand{\mybibhead}{\vspace{-1cm}}
\newcites{SM}{\mybibhead}

\usepackage{siunitx}
\DeclareSIUnit\torr{Torr}
\DeclareSIUnit\amagat{amg}

\usepackage{hyperref}

\newcommand{\IF}{\rho}
\newcommand{\QF}{\sigma}

\usepackage{graphicx}
\usepackage{bm}
\usepackage[mathlines]{lineno}
\usepackage{color}
\definecolor{mygreen}{rgb}{0,0.5,0}
\definecolor{mygrey}{rgb}{0.5,0.5,0.5}
\definecolor{myred}{rgb}{0.75,0,0}
\definecolor{myblue}{rgb}{0,0,0.75}
\definecolor{mymagenta}{cmyk}{0,1,0,0.12}
\definecolor{mycyan}{cmyk}{1,0,0,0.12}
\definecolor{myorange}{rgb}{1.,0.5,0}
\definecolor{myviolet}{rgb}{0.6,0.15,0.6}
\definecolor{mybrown}{cmyk}{0,0.50,1,0.41}

\usepackage{ulem}

\newcommand{\supzero}{^{(0)}}
\newcommand{\supone}{^{(1)}}

\newcommand{\supQ}{^{(Q)}}

\newcommand{\be}{\begin{equation}}
\newcommand{\ee}{\end{equation}}
\newcommand{\bea}{\begin{eqnarray}}
\newcommand{\eea}{\end{eqnarray}}

\newcommand{\bB}{{\bf B}}

\newcommand{\nn}{\nonumber \\ }

\newcommand{\nne}{\nonumber \\ & = &}
\newcommand{\nnequiv}{\nonumber \\ & \equiv &}

\newcommand{\nnp}{\nonumber \\ & & +}
\newcommand{\nnm}{\nonumber \\ & & -}

\newcommand{\bF}{{\bf F}}

\newcommand{\bT}{{\bf T}}
\newcommand{\bN}{{\bf N}}
\newcommand{\bX}{{\bf X}}
\newcommand{\NA}{{N_A}}
\newcommand{\Fmax}{{F_{\rm max}}}
\newcommand{\Pcyc}{{P_{+}}}
\newcommand{\Pbar}{{\bar{P}}}
\newcommand{\Bbar}{{\bar{B}}}
\newcommand{\supin}{^{({\rm in})}}

\newcommand{\ICFO}{ICFO - Institut de Ci\`encies Fot\`oniques, The Barcelona Institute of Science and Technology, 08860 Castelldefels (Barcelona), Spain}
\newcommand{\ICREA}{ICREA - Instituci\'{o} Catalana de Recerca i Estudis Avan{\c{c}}ats, 08010 Barcelona, Spain}
\newcommand{\HDU}{Department of Physics, Hangzhou Dianzi University, 310018, Hangzhou, China}

\begin{document}

\newcommand{\thetitle}{Sub pT/$\sqrt{\mathrm{Hz}}$ optical magnetometry with squeezed light}
\renewcommand{\thetitle}{Quantum enhancement of sensitivity and signal bandwidth in an optically-pumped magnetometer using squeezed light}
\renewcommand{\thetitle}{
Squeezed-light enhancement and backaction evasion in a high sensitivity optically-pumped magnetometer}

\title{\thetitle}

\author{C.\ Troullinou}
\affiliation{\ICFO}
\author{R. Jim\'enez-Mart\'{\i}nez}
\affiliation{\ICFO}
\author{J. Kong}
\affiliation{\HDU}
\author{V.\ G.\ Lucivero}
\affiliation{\ICFO}
\author{M. W. Mitchell}
\affiliation{\ICFO}
\affiliation{\ICREA}

\date{\today}

\begin{abstract}

We study the effect of optical polarization squeezing on the performance of a sensitive, quantum-noise-limited optically-pumped magnetometer. We use Bell-Bloom (BB) optical pumping to excite a \textsuperscript{87}Rb vapor containing \SI{8.2e12}{atoms\per\centi\meter\cubed} and Faraday rotation to detect spin precession.   The sub-\SI{}{\pico\tesla\per\sqrt\hertz} sensitivity is limited by spin projection noise (photon shot noise) at low (high) frequencies.  Probe polarization squeezing both improves high-frequency sensitivity and increases measurement
bandwidth, with no loss of sensitivity at any frequency, a direct demonstration of the evasion of measurement backaction noise. We provide a model for the quantum noise dynamics of the BB magnetometer, including spin projection noise, probe polarization noise, and measurement backaction effects. The theory shows how polarization squeezing reduces optical noise, while measurement backaction due to the accompanying ellipticity anti-squeezing is shunted into the unmeasured spin component. The method is compatible with high-density and multi-pass techniques that reach extreme sensitivity.

\end{abstract}

\maketitle


Optically-pumped magnetometers (OPMs) \cite{Budker2007}, in which an atomic spin ensemble is optically pumped \cite{HAPPER1972} and its spin-dynamics optically detected, are a paradigmatic quantum sensing technology with applications ranging from geophysics \cite{Dang2010} to medical diagnosis \cite{Boto2018} to searches for  physics beyond the standard model \cite{AbelPRL2020}.  OPMs are also a useful proving ground to test sensitivity enhancement techniques that may some day be applied to atomic clocks \cite{KnappeOL2005}, gyroscopes \cite{KornackPRL2005}, and co-magnetometers \cite{LeePRL2018, Limes2018}. In these sensors two quantum systems -- atoms and light -- interact to produce the signal.
Understanding and controlling the quantum noise in this interacting system is an outstanding challenge \cite{HuelgaPRL1997, AuzinshPRL2004, HorromPRA2012, NovikovaPRA2015}. 

At high atomic densities that give high OPM sensitivity, quantum noise of both atoms and light is important \cite{Budker2007}. Measurement backaction, including the effect of probe quantum noise on the spin system, becomes important in such conditions \cite{Vasilakis2011}, making it unclear whether squeezing of the probe light \cite{CavesPRD1981, GrangierPRL1987, Polzik1992, Han2016}, which reduces noise in one optical  component while increasing it in another, can reduce total noise in a high-sensitivity OPM. In contrast to squeezed-light enhancement in low-density OPMs \cite{WolfgrammPRL2010}, high-density squeezed-light OPMs \cite{Otterstrom2014} have to date shown a trade-off of sensitivity vs. quantum noise reduction \cite{Novikova2015, Zhang2021}, and a worsening of sensitivity due to probe squeezing  \cite{HorromPRA2012}.

To show that squeezing can indeed benefit a high-sensitivity OPM, we study a backaction evading measurement scheme based on  Bell-Bloom (BB) optical pumping \cite{Bell1961} and off-resonance probing. We model the quantum noise dynamics, including optical and spin quantum noise, and their interaction. We find that measurement backaction noise is shunted into a spin component that does not contribute to the signal. 
In this way the  scheme almost fully evades measurement backaction noise, including that associated with squeezing. We predict and experimentally demonstrate that squeezing improves the sensitivity of the OPM above the response bandwidth of the magnetometer, without significantly increasing noise in any part of the spectrum. Squeezing is also observed to improve the {measurement bandwidth} \cite{Shah2010}, i.e., the frequency range over which the sensitivity is within \SI{3}{\decibel} of its best value.

Our sensor achieves sub-\SI{}{\pico\tesla\per\sqrt\hertz} sensitivity to low-frequency finite fields, comparable to that of the best scalar OPMs implemented with mm-sized \cite{Gerginov2020} vapor cells and far better than previous squeezed-light enhanced OPMs \cite{WolfgrammPRL2010, HorromPRA2012, Otterstrom2014}.
The backaction evasion scheme is  compatible with sub-\SI{}{\femto\tesla\per\sqrt\hertz} methods including high-density \cite{Dang2010} and multi-pass \cite{ShengPRL2013} techniques, as well as with pulsed gradiometry \cite{Lucivero2021,Perry2020} and closed-loop \cite{LiPRAppl2020} techniques for operation at Earth's field \cite{Lucivero2019} and in unshielded environments \cite{Limes2020}. The BB technique also gives a clear view of the relationships among different noise sources. The results provide experimental input to the much-discussed question of whether squeezing techniques can, in practice, improve the performance of atomic sensors  \cite{HuelgaPRL1997,AuzinshPRL2004, HorromPRA2012, Zhang2021, MitchellRMP2020}.

\begin{figure*}[t]
	\centering
	\includegraphics[width=0.95\textwidth]{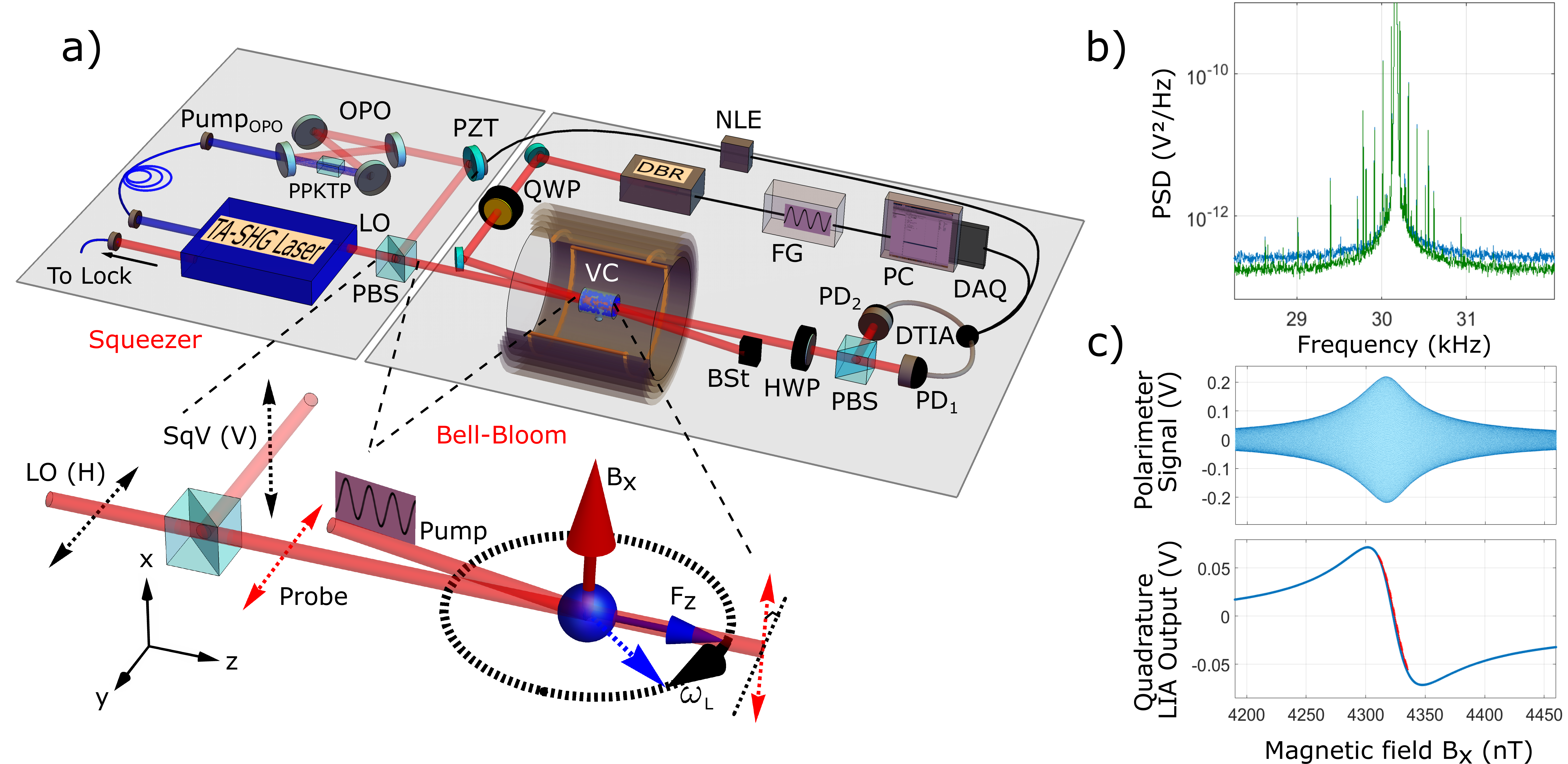}
	\caption{\textbf{Squeezed-light Bell-Bloom OPM. a) Experimental setup}. TA-SHG, Tapered Amplified Second Harmonic Generator; OPO, Optical Parametric Oscillator; PPKTP, Nonlinear crystal; LO, Local Oscillator; PBS, Polarizing Beam Splitter; QWP - Quarter Wave-plate; VC - Vapor Cell; BSt - Beam stopper; HWP - Half Wave-Plate; PD - Photodiode; DTIA - Differential Transimpedance Amplifier; DAQ - Data Acquisition; FG - Function Generator; NLE - Noise Lock Electronics. \textbf{``Bell-Bloom'' Inset}: Due to the magnetic field $B_x$ atomic spins precess at the Larmor frequency $\omega_L$ in the transverse plane. Synchronously modulated optical pumping maintains the atomic spin polarization. A linearly polarized cw probe undergoes paramagnetic Faraday rotation. \textbf{``Squeezer'' Inset:} Vertically-polarized squeezed vacuum is combined with horizontally-polarized LO on a polarizing beam splitter to generate a polarization squeezed probe. \textbf{b) Power Spectral Density (PSD).}  Power spectrum of the BB signal for coherent and squeezed-light around the Larmor frequency. The spectra are averages of 100 measurements, each one with duration of 0.5 sec. \textbf{c) Polarimeter signal (Top):} Signal $S_2$ (Eq.~(\ref{eq:S2Out})) for fixed pumping modulation frequency while scanning the magnetic field around resonance. \textbf{OPM signal (Bottom):} - lock-in quadrature output ($v$)  whose slope, calculated from the linear fit of the dispersive curve around the resonant field value, is used for the calibration of the magnetic sensitivity. 
	 }
	\label{fig:BBmagSetup}
\end{figure*}

\newcommand{\StwoNoise}{N_{S_2}}
\newcommand{\SthreeNoise}{N_{S_3}N_{S_3}}
\renewcommand{\StwoNoise}{\widetilde{S}_2}
\renewcommand{\SthreeNoise}{\widetilde{S}_3}

The experimental setup and coordinate system are shown in Fig.~\ref{fig:BBmagSetup}(a). Isotopically enriched $^{87}$Rb vapor and \SI{100}{\torr}  of N$_2$ buffer gas are contained in a cell with interior length \SI{3}{\centi\meter}. The cell, within a ceramic oven, is maintained by intermittent Joule heating at \SI{105}{\celsius} to create a $^{87}$Rb density of \SI{8.2e12}{atoms\per\centi\meter\cubed} and an optical transmission of about \SI{70}{\percent} for probe light blue detuned by \SI{20}{\giga\hertz} from the D$_1$ line. The cell and heater sit at the centre of four layers of cylindrical mu-metal shielding with cylindrical coils to control the bias field components $B_\alpha$ and gradients $\partial B_\alpha/\partial_z$, $\alpha \in \{x,y,z\}$.  A \SI{500}{\micro\watt} pump beam from a distributed Bragg reflector (DBR) laser, circularly polarized and current-tunable within the D$_1$ line at \SI{795}{\nano\meter}, propagates through the cell at a small angle from the $z$ axis. An extended cavity diode laser at \SI{795}{\nano\meter} is stabilized \SI{20}{\giga\hertz} to the blue from the $^{87}$Rb D$_1$ line with a fiber interferometer \cite{Kong2015} and frequency-doubled to produce violet light at \SI{397.4}{\nano\meter} (Toptica TA-SHG 110). The violet light is mode-cleaned in a polarization-maintaining fiber and then pumps a sub-threshold optical parametric oscillator to produce vertically-polarized squeezed vacuum at the laser fundamental frequency, as described in \cite{Predojevic2008}. The squeezed vacuum is combined on a polarizing beam-splitter with a mode-matched, horizontally-polarized ``local oscillator'' (LO) laser beam at \SI{795}{\nano\meter} to produce the polarization-squeezed probe. The relative phase between LO and squeezed vacuum is controlled by a piezo-electric actuator and active feedback using the broadband noise level of the signal as the system variable \cite{Predojevic2008}. In both coherent and squeezed-light probing, a \SI{400}{\micro\watt} beam is detected with a shot-noise-limited balanced polarimeter after the cell. The system is operated as a BB OPM at a finite field $\mathrm{B} =\SI{4.3}{\micro\tesla}$ by applying a low-noise current through the coils (current source Twinleaf CSUA300); gradients and other bias components are nulled. The DBR laser's current is square modulated with duty cycle 10$\%$ at angular frequency $\Omega = \omega_L \approx 2\pi \times \SI{30}{\kilo\hertz},$ equal to the angular Larmor frequency $\omega_L$.  The effect of the current modulation is to bring the laser frequency into optical resonance with the $F=1\rightarrow F'=1,2$ transitions once per modulation cycle. In each measurement cycle, the modulated pumping is maintained for \SI{0.5}{\second}. The resulting spin dynamics are observed as paramagnetic Faraday rotation of the probe beam. Under continuous, modulated pumping, the polarimeter signal oscillates with frequency $\Omega$, and shows noise from both spin projection noise and photon shot noise \cite{Lucivero2014}. 
The role of quantum noise can be qualitatively understood from a Bloch equation model described in detail in the Supplemental material \cite{QeBBMSOMFootnote}. 
The spins evolve according to the stochastic differential equation ${d\bF}/{dt} = {\bf V} + {\bf N}$, where ${\bf F}$ is the collective atomic spin vector, $\bN$ is a Langevin noise term and
\begin{eqnarray}
\label{eq:MeanSpinDynamics}
{\bf V} & = & -\gamma B \hat{x} \times \bF - \Gamma  \bF + P (\hat{z} \Fmax -\bF) 
\end{eqnarray}
is the drift rate \cite{QeBBMSOMFootnote}. Here $\gamma$ is the gyromagnetic ratio of $^{87}$Rb,  $\Gamma=1/T_2$ is the transverse relaxation rate,  $P$ is the optical pumping rate, $F_{\rm max} = \NA F$ is the maximum possible polarization, and $\NA$ is the atom number. Eq.  (\ref{eq:MeanSpinDynamics}) describes a spin oscillator with resonant frequency $\omega_L \equiv \gamma B = \gamma (B\supzero + B\supone)$, where $B\supzero$ is the time-average of $B$ and $|B\supone| \ll |B\supzero|$.

In the small-angle approximation appropriate here, the Faraday rotation signal can be written as \cite{KongNC2020} :
\begin{eqnarray}
\label{eq:S2Out}
S_2 = G S_1 F_z + N_{S_2},
\end{eqnarray}
where $S_\alpha$, $\alpha \in \{1,2,3\}$ indicate Stokes parameters at the output of the cell, $G$ is a coupling constant, and $N_{S_2}$ is the polarization noise of the detected Stokes component, a manifestation of quantum vacuum fluctuations \cite{CavesPRD1981}.

The oscillating spins and signal can be described in terms of slowly-varying quadratures $\IF, \QF, u, v$ via $F_z(t) = \IF \cos\Omega t + \QF \sin\Omega t$ and $S_2(t) =u \cos \Omega t +  v \sin \Omega t$. The in-phase ($u$) and quadrature ($v$) components are obtained by digital lock-in detection of the signal $S_2$. 
We set $\Omega=\gamma B\supzero$ to maximize $u$, at which point $v$ is linear in $B\supone$. Small changes in $B$ produce a linear change in the phase of the $S_2$ oscillation, such that $\widetilde{v}(\omega) = R(\omega) \widetilde{B}(\omega)$, where a tilde indicates a Fourier amplitude,
\bea
\label{eq:S2SignalQSlopeText}
R(\omega) & \equiv & 
 \frac{  \gamma \langle u \rangle}
{-i \omega  + \Delta\omega},
\eea
is the magnetic response, $\langle u \rangle = G S_1\supin \langle\IF\rangle$ is the signal amplitude, $\langle\IF\rangle$ is the equilibrium spin polarization, and $\Delta\omega \equiv \Gamma + \Pbar$ is the response bandwidth, where $\Pbar$ is the cycle-average of $P$. 
We compute the single-sided power spectral density of this signal, as ${\cal S}_v(\omega) \equiv |{\cal F}[N_v]|^2$, where ${\cal F}$ is the discrete Fourier transform implemented with a Hann window. 

The spin noise is
\begin{eqnarray} 
\label{eq:SpinNoiseInput}
{\bf N} & = & \bN_F + G S_3 \hat{z}  \times \bF,
\end{eqnarray}
where $\bN_F$ accounts for the noise introduced by pumping and relaxation, as required by the fluctuation-dissipation theorem \cite{QeBBMSOMFootnote}, 
$GS_3\hat{z}$ is the effective field produced by ac-Stark shifts \cite{HapperPR1967} due to the probe, and is effectively white.

\begin{figure}[t]
\centering
\includegraphics[width=\columnwidth]{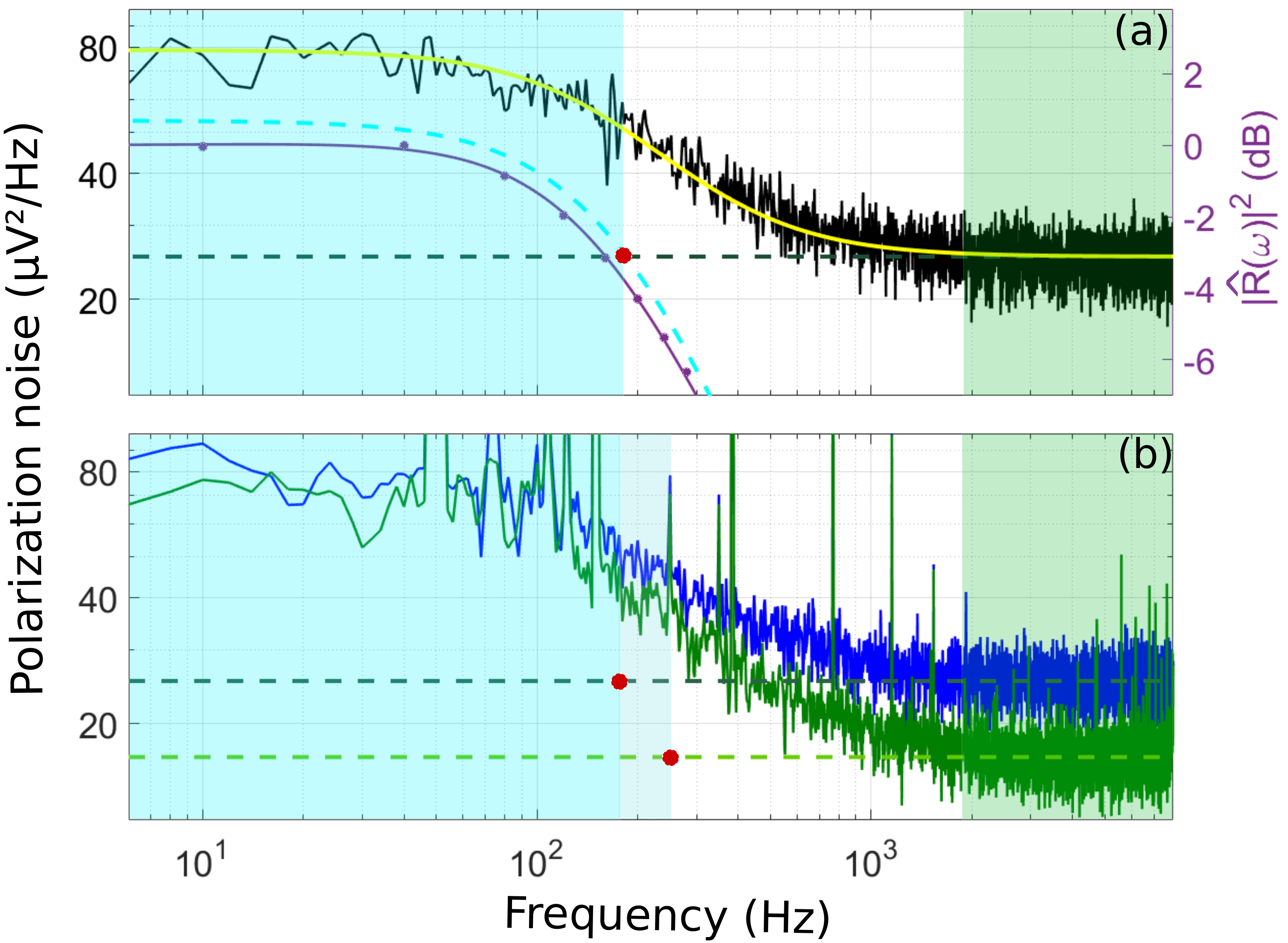}
\caption{
\textbf{Polarization rotation noise after demodulation.} \textbf{(a) Spin noise for unpolarized atoms.} We fit the spectrum (black) with a model function of Eq. (S45) to estimate the response bandwidth, the photon shot noise (dashed green line) and the low frequency spin projection noise. Subtracting the constant shot noise contribution from the fitted combined noise (yellow) we infer the spin projection noise curve in the unpolarized state probed (dashed cyan). These noise levels define the spin projection noise (cyan) and photon shot noise  (green) limited areas and the intermediate transition region (white). 
Purple dots and curve show, on the right axis, the measured normalized frequency response $|\hat{R}(\omega)|^2$ to an applied $B_x$ modulation, and its fit with Eq.~(\ref{eq:S2SignalQSlopeText}) with best fit parameter $\Delta\omega = \SI{170}{\hertz}$. 
\textbf{(b) Magnetometer noise for polarized atoms.} With \SI{500}{\micro\watt} of pump power, the noise spectrum of the magnetometer shows a very similar behaviour to the unpolarized spectrum and apart from the technical noise peaks at the power-line frequency and harmonics, quantum noise is dominant. At high frequencies, the noise level is reduced by 1.9 dB for squeezed-light (green), with respect to the coherent (blue) probing. The dashed lines and the red dots depict estimates of photon shot noise level and cross-over frequencies when the squeezer is on and off, respectively. 
}
\label{fig:FTvNoiseCohSNS}
\end{figure}

The three quantum noise sources affect differently the measurement. The azimuthal projection of $\bN_F$ contributes directly to the spin angle $\theta$, just as would a magnetic field, and thus with efficiency $\propto R(\omega)$. In contrast, $N_{S_2}$ is white noise, unrelated to the atomic response. Spectra of these two noise sources are shown in Fig.~\ref{fig:FTvNoiseCohSNS}(a) along with the experimentally measured magnetic response for comparison. The weak noise term $GS_3\hat{z}$ competes with the stronger $|B|\hat{x}$ in directing the spin precession, such that only its $\Omega$-resonant component has a first-order effect. Said effect only alters the $F_x$ component, which has no first order effect on the signal $S_2$. As a result, this BB magnetometer is backaction evading \cite{Shah2010, ColangeloN2017, ColangeloPRL2017}. Most importantly for the use of squeezed light, there is no deleterious effect from using squeezing to reduce the noise in $S_2$. While this necessarily increases the noise in $S_3$, said increase has no effect on the signal. Two potential benefits of optical squeezing are thus clear: it will reduce the noise for higher frequencies, and increase the frequency at which the noise $S_v(\omega)$ transitions from spin-noise dominated to photon shot-noise dominated. As we describe below, this improves both high-frequency sensitivity and measurement bandwidth of this quantum-noise-limited sensor. 

The calculation of magnetic sensitivity requires the above noise contributions to be normalized by the magnetic response. The latter is shown experimentally in Fig.~\ref{fig:FTvNoiseCohSNS}(a), and via the BB noise model \cite{QeBBMSOMFootnote} to have a characteristic roll-off described by a Lorentzian $\mathcal{L}(\omega)=(\Delta \omega)^2/(\omega^2+(\Delta \omega)^2)$ \cite{Shah2010}. 
The magnetic noise density is then 
\bea
\label{eq:Sensitivity}
{\cal S}_B(\omega) & = &   {\cal S}_v(\omega)|R(\omega)|^{-2} 
\nne \frac{\Delta \omega^2}{\gamma^2 \langle u \rangle^2} \left( {\cal S}_\sigma + \frac{1}{{\cal L}(\omega)} {\cal S}_{N_{S_2}} \right),
\eea
where ${\cal S}_{\sigma}$ and ${\cal S}_{N_{S_2}}$ are the noise spectral densities of the quadrature components of $F$ and $N_{S_2}$, respectively, and are frequency-independent. ${\cal S}_B(\omega)$ is nearly constant in the spin projection noise limited region and increases quadratically with frequency to double the low-frequency value at $\omega_{\SI{3}{\decibel}} \equiv \Delta \omega \sqrt{ {\cal S}_\sigma /{\cal S}_{N_{S_2}} +1}$. This frequency defines the \SI{3}{\decibel}  measurement bandwidth and grows with decreasing ${\cal S}_{N_{S_2}}$.

To demonstrate these advantages, we implement continuous-wave squeezed-light probing of the quantum-noise-limited BB OPM by using the experimental setup shown in Fig.~\ref{fig:BBmagSetup}(a). As already described, the resulting optical beam is horizontally polarized with squeezed fluctuations in the diagonal basis, i.e., squeezed in $S_2$. For an OPO pump power of \SI{40.6}{\milli\watt} the generated polarization squeezing is at \SI{2.4}{\decibel} before the cell, as measured from the PSD of the signal from an auxiliary balanced polarimeter. Because of \SI{30}{\percent} absorption losses, \SI{1.9}{\decibel} of squeezing is observed in the PSD of both the BB polarimeter signal, shown in Fig.~\ref{fig:BBmagSetup}(b), and the demodulated quadrature component, shown in  Fig.~\ref{fig:FTvNoiseCohSNS}(b). 

We compute the experimental sensitivity following prior work on BB magnetometers  \cite{Jimenez-Martinez2012,Gerginov2017,Gerginov2020}, as 
\bea
\label{eq:SensitivityMeasured}
{\cal S}_B(\omega) & = &  \left( \frac{dv}{dB}\right)^{-2} \frac{{{\cal S}_v(\omega)}}{|\hat{R}(\omega)|^2},
\eea
where ${\cal S}_v(\omega)$ is the observed noise in the lock-in quadrature component $v$, $dv/dB$ is the slope of the  quadrature signal and  $|\hat{R}(\omega)|^2  \equiv |R(\omega)/R(0)|^2$ is the normalized frequency response of the spins to a modulation of the field $B_x$, shown in Figs. ~\ref{fig:FTvNoiseCohSNS}(b),~\ref{fig:BBmagSetup}(c) and \ref{fig:FTvNoiseCohSNS}(a) respectively. Measurement of the magnetometer frequency response to a fixed amplitude sine wave magnetic field modulation in the range of \SI{10}{\hertz} to \SI{2.4}{
\kilo\hertz} is used to experimentally determine $|R(\omega)|^2$ \cite{QeBBMSOMFootnote}.
\begin{figure}[t]
	\centering
	\includegraphics[width=0.5\textwidth]{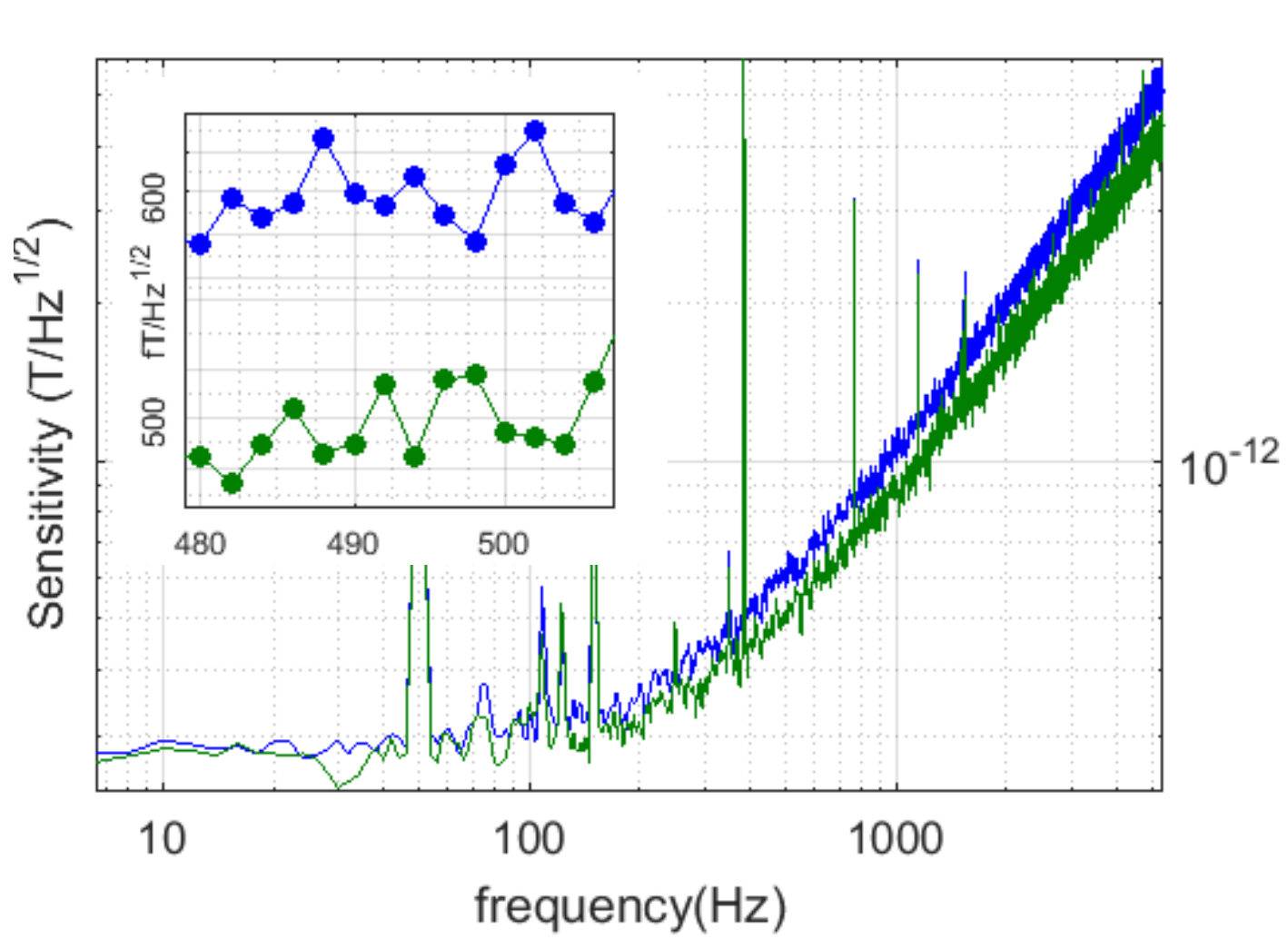}
	\caption{\textbf{Magnetic Sensitivity.} Sensitivity spectra for BB magnetometer probed with coherent (blue) and squeezed light (green). All data acquired with $P_{\rm probe} = \SI{400}{\micro\watt}$, $P_{\rm pump} = \SI{500}{\micro\watt}$, $T = \SI{105}{\celsius}$, $n = \SI{8.2e12}{atoms\per\centi\meter\cubed}$, $B_0 =\SI{4.3}{\micro\tesla}$, $f_{\rm mod} = \SI{30.164}{\kilo\hertz}$. Polarization squeezing was \SI{2.4}{\decibel} before the atomic cell and \SI{1.9}{\decibel} at the detectors. }   
	\label{fig:SensitivitySqMAGT109C} 
\end{figure}

\newcommand{\supSQL}{^\mathrm{SQL}}
\newcommand{\supSQ}{^\mathrm{sq}}
\newcommand{\subthreedB}{_{\SI{3}{\decibel}}}
\newcommand{\noiseratio}{\zeta}

Turning on the squeezer causes  ${\cal S}_{N_{S_2}}$ to drop to $\xi^2$ times its coherent-state value ${\cal S}_{N_{S_2}}\supSQL$, where $\xi^2$ is the squeezing parameter \cite{Lucivero2016, Lucivero2017a}. 
The predicted magnetic power spectral density is then 
\bea
{\cal S}_B(\omega) & = & \frac{\Delta \omega^2}{\gamma^2 \langle u \rangle^2} [ {\cal S}_\sigma + \frac{\xi^2}{{\cal L}(\omega)} {\cal S}_{N_{S_2}}\supSQL ]
\nne {\cal S}_B\supSQL(\omega) \frac{1+{\cal L}(\omega)\xi^2 \noiseratio^{-2} }{1+{\cal L}(\omega) \noiseratio^{-2}},
\eea
where $\noiseratio^2 \equiv {{\cal S}_\sigma}/{\cal S}_{N_{S_{2}}}\supSQL$.

The enhancement due to squeezing is evident in the high frequency part of the experimental spectrum, shown in Fig.~\ref{fig:SensitivitySqMAGT109C}. At the detection frequency of \SI{490}{\hertz}, a polarization squeezing of \SI{1.9}{\decibel} results in a \SI{17}{\percent} quantum enhancement of magnetic sensitivity, from \SI{600}{\femto\tesla\per\sqrt\hertz} down to \SI{500}{\femto\tesla\per\sqrt\hertz}. As seen in Fig.~\ref{fig:SensitivitySqMAGT109C}, squeezing does not add noise to any region of the spectrum. This is a direct experimental demonstration that the BB technique evades backaction associated with the anti-squeezed $S_3$ component.
 
Squeezed-light probing also increases the \SI{3}{\decibel} measurement bandwidth \cite{Shah2010}. For the data presented in Fig.~\ref{fig:SensitivitySqMAGT109C}, the original measurement bandwidth of \SI{275}{\hertz} is already higher than the response bandwidth $\Delta \omega =  \SI{170}{\hertz}$ and it is further increased to \SI{320}{\hertz}, with about \SI{15}{\percent} of quantum enhancement. This result agrees with the predicted improved \SI{3}{\decibel} measurement bandwidth estimated via
\bea
\omega\supSQ\subthreedB =  \omega\supSQL\subthreedB \sqrt{\frac{1+\noiseratio^2\xi^{-2}}{1+\noiseratio^2}}
\eea
The quantum advantages demonstrated here are limited by the  squeezing produced by our OPO \cite{Predojevic2008}, and by probe transmission losses. 
Optical losses for the probe can in principle be made arbitrarily small without altering the other characteristics of the magnetometer, by increasing the probe detuning while boosting the probe power to keep constant the probe power broadening. More recent OPO designs \cite{HanOE2016, VahlbruchPRL2016} have demonstrated up to \SI{15}{\decibel} of squeezing. 

In conclusion, we have demonstrated that a polarization-squeezed probe can give both higher sensitivity and larger measurement bandwidth in a sensitive optically pumped magnetometer. In contrast to squeezed-light probing of optomechanical sensors such as gravitational wave detectors \cite{McCullerPRL2020}, the sensitivity advantage at high frequencies comes without the cost of increased backaction noise at low frequencies. This occurs because QND measurement of a precessing spin system shunts backaction effects into the unmeasured spin degree of freedom \cite{ColangeloN2017}, something not possible in a canonical system such as a mechanical oscillator \cite{HePRA2011}. Squeezed-light probing is compatible with and complementary to other methods to enhance sensitivity and bandwidth, including spin-exchange relaxation suppression \cite{SavukovBook2017}, pulsed geometries \cite{Lucivero2019, Limes2020}, multi-pass geometries \cite{ShengPRL2013}, Kalman filtering \cite{Jimenez-Martinez2018} and closed-loop techniques \cite{LiPRAppl2020}.

\section{}

Acknowledgements: We thank Michele Gozzelino and Dominic Hunter for laboratory assistance and helpful discussions and Vindhiya Prakash for feedback on the manuscript.
This project was supported by 
H2020 Future and Emerging Technologies Quantum Technologies Flagship projects MACQSIMAL (Grant Agreement No. 820393) and  QRANGE (Grant Agreement No.  820405); 
H2020 Marie Sk{\l}odowska-Curie Actions projects ITN ZULF-NMR  (Grant Agreement No. 766402) and PROBIST (Grant Agreement No. 754510), Spanish Ministry of Science projects OCARINA (Grant No. PGC2018-097056-B-I00 FEDER ``A way to make Europe'') and ``Severo Ochoa'' Center of Excellence CEX2019-000910-S
Generalitat de Catalunya through the CERCA program; 
Ag\`{e}ncia de Gesti\'{o} d'Ajuts Universitaris i de Recerca Grant No. 2017-SGR-1354;  Secretaria d'Universitats i Recerca del Departament d'Empresa i Coneixement de la Generalitat de Catalunya, co-funded by the European Union Regional Development Fund within the ERDF Operational Program of Catalunya (project QuantumCat, ref. 001-P-001644); Fundaci\'{o} Privada Cellex; Fundaci\'{o} Mir-Puig,  J. K. acknowledges the support from NSFC through Grant No. 11935012, 12005049.

\bibliographystyle{./biblio/apsrev4-1}
\bibliography{./biblio/PUMPINGBELLBLOOM}

\newpage

\begin{widetext}
\newpage
\noindent
{\Large Supplemental material for: \textit{\thetitle}}
\end{widetext} 

\setcounter{page}{1}
\renewcommand{\thepage}{S\arabic{page}} 
\renewcommand{\thesection}{S\arabic{section}}  
\renewcommand{\thetable}{S\arabic{table}}  
\renewcommand{\thefigure}{S\arabic{figure}}
\renewcommand{\theequation}{S\arabic{equation}} 
\setcounter{figure}{0}

\section{Lock-in detection }

Lock-in detection is implemented offline in Matlab/Octave: The digitized signal from the balanced polarimeter is demodulated by multiplying by $\cos(\omega_{\rm mod} t + \phi)$ (in-phase component) or $\cos(\omega_{\rm mod} t + \phi + \pi/2)$ (quadrature component), where $\omega_{\rm mod}$ and $\phi$ are the angular frequency and phase, respectively, of the square waveform used to modulate the pump laser current. 

\section{Magnetometer responsivity }
\begin{figure}[t]
	\centering
	\includegraphics[width=\columnwidth]{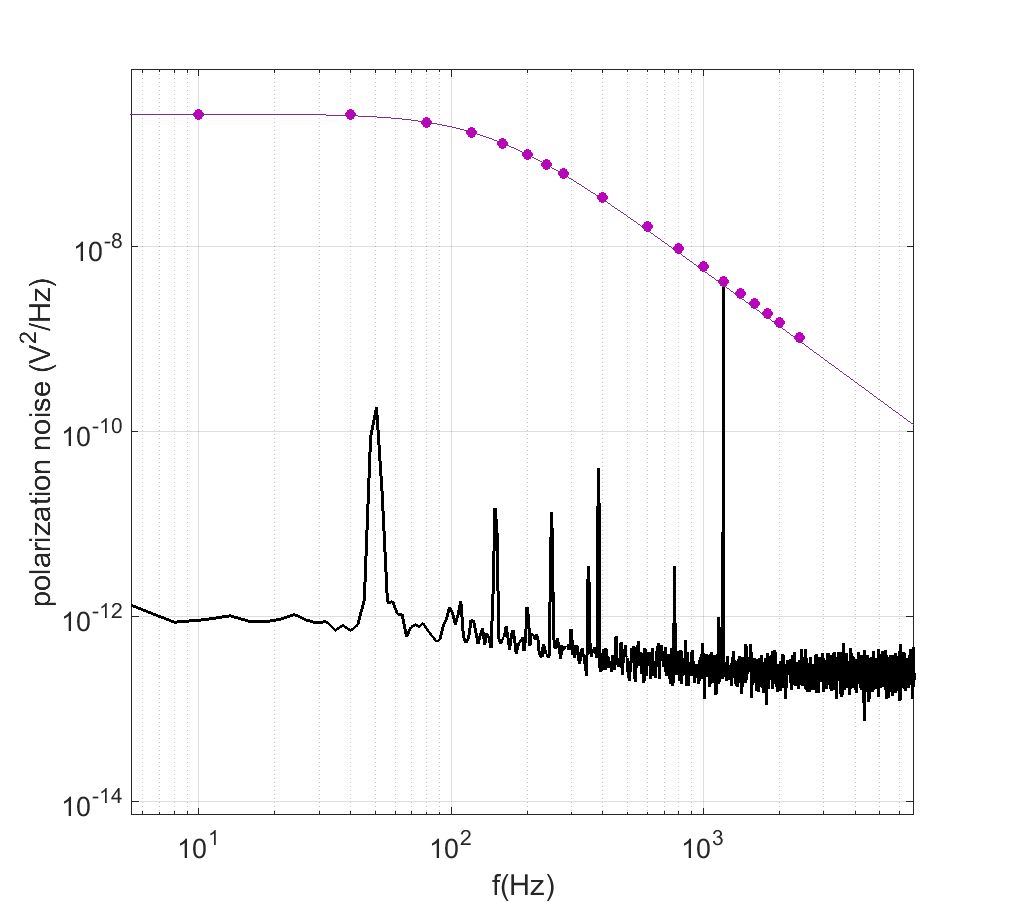}
	\caption{\textbf{Experimentally measured magnetic response.} The black spectrum is the polarization noise of the quadrature signal from the LIA output when the frequency of the magnetic field modulation is set at \SI{1.2}{\kilo\hertz}. The spectrum is averaged for 20 measurement cycles of \SI{0.5}{\second} each. The purple dots show the peak of PSD resonance when varying the magnetic modulation frequency between \SI{10}{\hertz} and \SI{2.4}{\kilo\hertz}. Purple line shows fit with a Lorentzian centered at zero frequency. }
	\label{fig:expMagResp} 
\end{figure}

In order to characterize the magnetic response of our sensor, we operate the magnetometer in the conditions described in the text while imposing on top of the $B_x = \SI{4.3}{\micro\tesla}$ dc field component a sinusoidal modulation of amplitude \SI{0.36}{\nano\tesla}. This magnetic modulation appears as a peak in the spectrum of the magnetometer signal, proportional to $|R(\omega_{\rm mod})|^2$ from Eq.~(\ref{eq:S2SignalQSlopeText}), as shown in Fig.~\ref{fig:expMagResp}. In this way we measure $|R(\omega_{\rm mod})|^2$ (dots in Fig.~\ref{fig:expMagResp}). We fit with a Lorentzian centered at zero frequency to obtain the parameter $\Gamma + \Pbar  = \SI{170}{\hertz}$ used in the calculation of magnetic sensitivity and for comparison with the spin projection noise curve of  Fig.~\ref{fig:FTvNoiseCohSNS}(a). 

\newcommand{\subeff}{_{\rm eff}}
\newcommand{\subint}{_{\rm int}}
\renewcommand{\subint}{_{\rm spin}}
\newcommand{\subhfs}{_{\rm hfs}}
\newcommand{\subcycle}{_{\rm cyc}}

\section{Quantum noise theory of the Bell-Bloom magnetometer}

Here we describe a quantum noise model for the Bell-Bloom magnetometer. To keep the model reasonably simple while still retaining the essential physics, the model describes only a single spin population. In the experimental atomic system, operated outside of the SERF regime, the two hyperfine components precess with equal but opposite gyromagnetic ratios, contribute differently to the probe polarization rotation, and are resonant with the pump laser at different times in the pump laser's modulation cycle. Due to these differences, we do not expect to obtain quantitatively accurate  sensitivity curves from this model. Nonetheless, we see that the model reproduces very accurately the shape of the observed BB noise and sensitivity curves, including the effect of squeezed light on sensitivity and measurement bandwidth. This gives us reason to believe that the model captures the essential physics of quantum noise in the BB magnetometer.

The model employs Bloch equations to describe the spin dynamics, in which the spin evolves as \bea
\label{eq:OriginalSpinDynamics}
\frac{d}{dt} \bF & = & (-\gamma \bB + G S_3 \hat{z} ) \times \bF - \Gamma  \bF + P (\hat{z} \Fmax -\bF) \nnp \bN_F, 
\eea
where the first term describes precession under the magnetic field $\bB$ and optically-induced effective field $G S_3 \hat{z}$.  $\Fmax = \NA F$ is the maximum possible polarization, and $F$ is the spin quantum number.  The time-dependent optical pumping rate is $P$, and non-pumping relaxation, including spin relaxation due to atomic effects, i.e. spin-exchange and spin-destruction collisions, atomic diffusion and relaxation due to probing, i.e., power broadening, is described by the rate $\Gamma$. 

$\bN_F$ is a Langevin term, which accounts for noise introduced by both kinds of relaxation. It is Gaussian white noise with covariance \bea
\label{eq:NAutocorr}
\langle N_{F_i}(t)N_{F_j}(t') \rangle &=& 2 \frac{F(F+1)}{3} \NA [\Gamma + P(t)] \delta_{ij} \delta(t-t')
\nnequiv G_{N_F}  \delta_{ij} \delta(t-t').
\eea
as shown in the below Section ``Diffusion term.''

The power spectral density is the Fourier transform of this correlation function \bea
{\cal S}_{F_i}(f) &=&  \int dt'  \, \langle N_{F_i}(t)N_{F_j}(t') \rangle e^{-i2\pi ft'} = G_{N_F} \delta_{ij}. 
\hspace{9mm}
\eea

\section{Perturbative approach}

Eq.~(\ref{eq:OriginalSpinDynamics}) is not easy to solve exactly: it contains terms like $\gamma \bB \times \bF$ that are products of the time-dependent $\bB$ (what we are trying to measure) and the time-dependent $\bF$ (the spin that responds to it). In the scenario of interest, the time dependent quantities divide into strong, predictable ones and weak, noisy or to-be-measured ones.  This motivates a perturbative treatment, in which we write $\bB = \bB\supzero + \alpha \bB\supone$ where $\bB\supone$ is a small unknown perturbation on top of the strong, known $\bB\supzero$ and $\alpha$ is a perturbation parameter that we take to unity at the end. Similarly, we write $S_3 = \alpha S_3\supone$ and $\bN_F = \alpha \bN_F\supone$, and we expand $\bF$, which depends on the preceding variables, as $\bF = \bF\supzero + \alpha \bF\supone + \ldots $. 
 Eq.~(\ref{eq:OriginalSpinDynamics}) now becomes
\begin{widetext}
\bea
\label{eq:ExpandedSpinDynamics}
\frac{d}{dt} (\bF\supzero + \alpha \bF\supone + \ldots) & = & -\gamma (\bB\supzero + \alpha \bB\supone)  \times (\bF\supzero + \alpha \bF\supone+ \ldots)  +   \alpha G S_3\supone \hat{z} {\times} (\bF\supzero + \alpha \bF\supone + \ldots) \nnm \Gamma (\bF\supzero + \alpha \bF\supone+ \ldots) + P [\hat{z} \Fmax -(F_z\supzero + \alpha F_z\supone + \ldots)] + \alpha \bN_F\supone \hspace{6mm}
\eea
\end{widetext}
We can formally solve the zero-th order case, i.e. with $\alpha = 0$. The solution is the $\bF\supzero$ that satisfies
\bea
\label{eq:ZeroOrderDynamics1}
\frac{d}{dt} \bF\supzero & = & -\gamma \bB\supzero  \times \bF\supzero  - \Gamma  \bF\supzero \nnp P (\hat{z} \Fmax -\bF\supzero) \hspace{6mm} 
\eea
or \bea
\label{eq:ZeroOrderDynamics2}
\left( \frac{d}{dt}  + \gamma \bB\supzero  \times  + \Gamma + P \right) \bF\supzero & = & P \hat{z} \Fmax.
\eea

The solution of this ordinary differential equation will simply be a function of time, not a stochastic process. It can be further used in the first order, i.e. $O(\alpha)^1$, dynamics, which is described by

\bea
\label{eq:FirstOrderDynamics}
\frac{d}{dt} \bF\supone & = & -\gamma ( \bB\supone  \times \bF\supzero +  \bB\supzero  \times \bF\supone)  + G S_3 \hat{z}  \times \bF\supzero \nnm \Gamma  \bF\supone - P \hat{z} \bF\supone +  \bN_F\supone \hspace{6mm}
\eea
or
\bea
\label{eq:FirstOrderDynamics2}
\left( \frac{d}{dt}  + \gamma \bB\supzero  \times + \Gamma + P \right) \bF\supone & = - & \gamma  \bB\supone  \times \bF\supzero +  \bN_F\supone \nnp  G S_3 \hat{z}  \times \bF\supzero.  \hspace{6mm} \hspace{6mm}
\eea
Note that this equation is linear in the unknown $\bF\supone$. 

The input light is polarized so that the Stokes parameter $\langle S_1 \rangle$ is maximum, and the readout Stokes parameter $S_2$ is given by
\bea
\label{eq:S2Signal}
S_2 & = &  G S_1 F_z + N_{S_2}
\eea
where $N_{S_2}$ is the quantum noise in that polarization component.  We note that $S_3$ is on average zero, but will have fluctuations that we describe by another Langevin term $N_{S_3}$, which has an uncertainty relation with respect to $N_{S_2}$.

We thus have five noise components represented here:  three components of $\bN_F$, plus $N_{S_2}$ and $N_{S_3}$.  We do not expect any of these to enter the signal in exactly the same way. We note, for example, that one component of $\bN_F$ (and of $\bF$) is out of the plane of precession,  i.e., in the direction of $\bB$, and does not contribute to the signal.  Another component of $\bN_F$ will be the radial component, i.e., parallel to $\bF$, and thus contributing noise to  the amplitude of the signal, while the remaining component of $\bN_F$ will be the azimuthal component, i.e., normal to both $\bB$ and $\bF$, and thus adding noise to the angle of precession.  This last one we can expect to look like magnetic signal and thus to be the most relevant atomic noise.  As regards the optical noises, $N_{S_2}$ will be a white noise that directly enters the measurement record, while $N_{S_3}$ will enter only to the degree that it can perturb the atomic spin precession.

\section{Bell-Bloom scenario - harmonic drive}
We now specialize to the BB scenario.  We take $\bB\supzero = |\bB\supzero| \hat{x}$ (a constant), and we assume that the pumping $P(t)$ is periodic and close to resonance with the Larmor precession, with an amplitude that is constant over time.  Without loss of generality we choose the time origin such that $\int_0^{2\pi/\Omega} P(t) \exp [i \Omega t] dt$ is positive real, meaning that $P(t)$ is in some sense centered on $\Omega t = 0, 2\pi, 4\pi, \ldots$. Because $\bB\supzero$ is along the $\hat{x}$ direction and the pumping is in the $\hat{z}$ direction, $\bF\supzero$ will be in the $y$--$z$ plane.  Remembering that the measured component is $F_z$, we can now see that the $S_3$ term will make no zero-th order or first-order contribution to the signal.   From Eq. (\ref{eq:FirstOrderDynamics}), its contribution to $\bF\supone$ is  $\propto \hat{z} \times \bF\supzero$ which is along the $\hat{x}$ direction, i.e. the component of $\bF\supone$ that is not measured.  For this reason, we can drop this term from here on.

Similarly, the different components of  $\bB\supone$, the perturbation to the field, have a different effects:  The $\hat{x}$ component increases the magnitude of $\bB$ in first order, and thus will change the precession rate.  The $\hat{y}$ and $\hat{z}$ components will (in first order) only tip the axis of the precession, which introduces in first order an oscillating $F_x$ component.   But again, since $F_x$ is not measured, this has no measurable first-order effect. All of which is summarized in the statement that the BB OPM is a scalar magnetometer, sensitive only  the magnitude of the field, which in first order involves only the bias $\bB\supzero$ and the component of $\bB\supone$ along $\bB\supzero$.  For this reason, we will consider from here on only $B_x\supone$.

\subsection{Rotating frame}
We assume the system has a resonance that is of reasonably high Q factor, which is to say that relaxation and pumping effects are not strong during one cycle, and forces that are resonant can accumulate over several cycles.  It is usual in such scenarios to describe the dynamics in a rotating frame, and to apply the rotating wave approximation. Here we define the frame rotating at $\Omega$, the angular frequency of the drive.  Given a vector  $\bX(t)$, the rotating-frame expression of $\bX$ is $X_+ = (i X_y + X_z) \exp[i \Omega t]$. It will also be useful to have the cycle-averaged version
\bea
\langle X_+\rangle\subcycle(t) & \equiv &  \frac{\Omega}{2\pi} \int_{t-\pi/\Omega}^{t + \pi / \Omega}[i X_y(t') + X_z(t')] \exp[i \Omega t'] \, dt'
\nn
\eea

The spin itself, for example, is described by $F_+ = (i F_y + F_z) \exp[i \Omega t]$, with the consequence that
\bea
F_z & = & {\cal R}[F_+ e^{-i \Omega t} ] = {\cal R}[F_+] \cos \Omega t +{\cal I}[F_+] \sin \Omega t  \hspace{6mm}\\
F_y & = & {\cal I}[F_+ e^{-i \Omega t} ] = - {\cal R}[F_+] \sin \Omega t  + {\cal I}[F_+] \cos \Omega t \hspace{6mm}
\eea
We note that in this representation
\bea
\hat{x} \times (\hat{y} F_y + \hat{z} F_z) & = & -\hat{y} F_z + \hat{z} F_y
\eea
 is accomplished by $F_z \rightarrow F_y$ and $F_y \rightarrow -F_z$, which is the same as $F_+  \rightarrow -i F_+ $.  From the product rule
 \bea
 \frac{d}{dt} [F_+ e^{-i \Omega t}] &=&  \left( \frac{d}{dt} {F}_+  - i  \Omega F_+\right)  e^{-i \Omega t}
 \eea
 we see that the time derivative in the rotating frame is given by $ {d}/{dt}  \rightarrow  {d}/{dt}  - i  \Omega $.

The torque produced by a field $\bB$ along the $\hat{x}$ direction can be written
\bea
\bT &\equiv&  \gamma \bB \times \bF
\nne \gamma |B| \hat{x} \times \{ \hat{y} {\cal I}(F_+ \exp[-i\Omega t] ) + \hat{z} {\cal R}(F_+ \exp[-i\Omega t] )\}
\nne \gamma |B| \{ \hat{z} {\cal I}(F_+ \exp[-i\Omega t] ) - \hat{y} {\cal R}(F_+ \exp[-i\Omega t] )\}
\nne \gamma |B| \left\{
\hat{z} ( {\cal I}[F_+]\cos\Omega t - {\cal R}[F_+]\sin\Omega t)
\right. \nonumber \\
& & \left. - \hat{y} ( {\cal R}[F_+]\cos\Omega t + {\cal I}[F_+]\sin\Omega t)
\right\}
\eea
so that 

\bea
\label{eq:BHarmonic2}
\langle T_{+}\rangle\subcycle(t) &=&   \frac{\Omega}{2\pi} \int_{t-\pi/\Omega}^{t + \pi / \Omega}  [ i T_{y}(t')  + T_{z}(t') ]e^{i \Omega t'} dt' \nne  \gamma |\Bbar_x|  \frac{1}{2}( {\cal I}[F_+] - i{\cal R}[F_+]  -i{\cal R}[F_+] + {\cal I}[F_+] )
\nne  \gamma |\Bbar| ({\cal I}[F_+] -i{\cal R}[F_+] )
\nne  \gamma |\Bbar| (- i F_+)
\eea
where $\Bbar =  \frac{\Omega}{2\pi} \int_{t-\pi/\Omega}^{t + \pi / \Omega}   B(t')  \, dt'$ is the cycle-averaged field strength.  Eq. (\ref{eq:BHarmonic2}) fits our expectations; it describes a contribution to the precession rate. 

\subsection{Demodulation}

Supposing we have a signal $Y_+$, which is in general complex. In the ``lab frame,'' this is
\bea
\label{eq:YSignal}
Y & = &   {\cal R}[Y_+]\cos \Omega t +  {\cal I}[Y_+] \sin \Omega t
\eea
the in-phase and quadrature components are just the pre-factors of the $\cos \Omega t$ and $\sin \Omega t$ terms, respectively i.e. ${\cal R}[Y_+]$  and ${\cal I}[Y_+]$, respectively.

\subsection{Order zero}

We want the steady-state solution to Eq. (\ref{eq:ZeroOrderDynamics2}), i.e. with $dF_+/dt \rightarrow 0$.  In addition to the above representations of $d/dt$ etc., we need to represent the pump.  Again, the cycle-averaged value is appropriate, and simpler than the general form.  We use this to define an effective pump rate $\Pcyc$:
\bea
\label{eq:PHarmonic}
\langle P_+\rangle\subcycle &=&   \frac{\Omega}{2\pi} \int_{t-\pi/\Omega}^{t + \pi / \Omega}   P(t') e^{i \Omega t'} \, dt' \equiv \Pcyc.
\eea
  Without loss of generality, we will assume that $\Pcyc$ is real, meaning that the pump is in some sense ``centred'' on the points $\Omega t = 0, 2\pi, 4\pi, \ldots$.  We note that while $\Pcyc$ is relevant for describing the forcing effect of the drive, the relaxation effects, including decay rate and noise, concern rather the cycle-averaged mean pump
\bea
\label{eq:PAve}
\langle P\rangle\subcycle &=&   \frac{\Omega}{2\pi} \int_{t-\pi/\Omega}^{t + \pi / \Omega}   P(t')  \, dt' \equiv \Pbar.
\eea

Using these rotating-frame representations, we transform Eq. (\ref{eq:ZeroOrderDynamics2}) into a simple algebraic equation:
\bea
\label{eq:ZeroOrderSteadyRot}
( - i  \Omega - i \gamma |B\supzero|   + \Gamma + \Pbar) F_+\supzero  & = &  \Pcyc \Fmax,
\eea
with solution
\bea
\label{eq:ZeroOrderSteadyRot2}
F_+\supzero  & = &  \frac{\Pcyc \Fmax}{- i  \Omega - i \gamma |B\supzero|   + \Gamma + \Pbar}  .
\eea
\newcommand{\omegaLo}{\omega_L\supzero}
We note that $\gamma < 0$, and that $-\gamma |B\supzero| \equiv \omegaLo$, the Larmor (angular) frequency to zero order, so that
\bea
\label{eq:ZeroOrderSteadyRot3}
F_+\supzero  & = &  \frac{\Pcyc \Fmax}{i (\omegaLo-  \Omega )  + \Gamma + \Pbar}  .
\eea
This describes a Lorentzian resonance that saturates, with the effective linewidth  $\Gamma + \Pbar$ increasing in such a way that the polarization $\Fmax \Pcyc/\Pbar \le \Fmax$ is never exceeded.

\subsection{Order one}

We now translate Eq. (\ref{eq:FirstOrderDynamics2}) to the rotating-frame picture. This time, we must keep the time derivative term $dF_+/dt$.  We find
\bea
\label{eq:FirstOrderDynamicsRot}
\left( \frac{d}{dt} + i (\omegaLo-\Omega)   + \Gamma + \Pbar \right) F_+\supone & = &  i \gamma  \Bbar\supone F_+\supzero +  N_{F_+}.  \nonumber \\   \hspace{6mm}
\eea
This is a Langevin equation, because of the noise term $N_{F_+}$, and also it contains the unknown $B\supone$, which depends on time. The equation has real and imaginary parts, which are coupled.  In the demodulation, only the imaginary part, corresponding to the quadrature component, is used to infer the field, so we separate  the real and imaginary parts
\begin{widetext}
\bea
\label{eq:FirstOrderDynamicsRotReIm}
\left(
\begin{array}{cc}
 \frac{d}{dt} + \Gamma + \Pbar & -(\omegaLo-\Omega) \\
\omegaLo-\Omega &   \frac{d}{dt} + \Gamma + \Pbar
\end{array}
\right)
\left( \begin{array}{c} {\cal R}[F_+\supone] \\ {\cal I}[F_+\supone] \end{array} \right) & = &
\left( \begin{array}{c} {\cal R}[N_{F_+}] \\  \gamma \Bbar\supone F_+\supzero + {\cal I}[N_{F_+}] \end{array} \right)
\eea
 We can solve this in the Fourier domain taking $\omega$ as the frequency variable of the Fourier spectrum (note that this is different from $\omega_L$ and from $\Omega$, which are constants), $d/dt \rightarrow -i\omega$, at which point we have
\bea
\label{eq:FirstOrderDynamicsRotReIm2}
\left(
\begin{array}{cc}
-i\omega + \Gamma + \Pbar & -(\omegaLo-\Omega) \\
\omegaLo-\Omega &   -i\omega + \Gamma + \Pbar
\end{array}
\right)
\left( \begin{array}{c} {\cal R}[F_+\supone](\omega) \\ {\cal I}[F_+\supone](\omega) \end{array} \right) & = &
\left( \begin{array}{c} {\cal R}[N_{F_+}](\omega) \\ \gamma \Bbar\supone(\omega) F_+\supzero + {\cal I}[N_{F_+}](\omega) \end{array} \right)
\eea
with solution
\bea
{\cal I}[F_+\supone](\omega) & = & \frac{[{\cal I}[N_{F_+}](\omega) + \gamma \Bbar\supone(\omega) F_+\supzero] (-i\omega + \Gamma + \Pbar) - (\omegaLo - \Omega){\cal R}[N_{F_+}](\omega) }
{[-i(\omega -\omegaLo + \Omega) + \Gamma + \Pbar] [ -i (\omega + \omegaLo - \Omega) + \Gamma + \Pbar] }
\eea
\end{widetext}

We see now that the response of $F_+$ to the field perturbation $B\supone(\omega)$ and to spin noise ${\cal I}[N_{F_+}](\omega)$ have the same frequency dependence, because they enter in exactly the same way.  Moreover, ${\cal I}[N_{F_+}](\omega)$ is constant, i.e. the noise is white.  This is the basis for saying that the spin noise and response to the magnetic field are matched \cite{Shah2010}.

\subsection{Resonant case}
We now specialize to the case of resonant excitation, i.e., $\Omega = \omegaLo$, which gives maximum signal and is the natural operating point for the BB magnetometer. We find 
\bea
\label{eq:ZeroOrderSteadyRes}
F_+\supzero  & = &  \frac{\Pcyc \Fmax}{\Gamma + \Pbar}
\eea
and
\bea
\label{eq:FirstOrderDynamicsRes}
{\cal I}[F_+\supone](\omega) & = & \frac{{\cal I}[N_{F_+}](\omega) + \gamma \Bbar\supone(\omega) F_+\supzero  }
{-i\omega + \Gamma + \Pbar }
\eea

\subsection{Optical signal, responsivity and sensitivity}
The optical signal $S_2$ is demodulated to obtain the quadrature component $S_{2}\supQ$. Writing this in the frequency domain we have
\bea
\label{eq:S2Signal2}
S_{2}\supQ(\omega) & = & {\cal I}[N_{S_2+}](\omega) + G S_1 {\cal I}[F_+\supone](\omega)
\eea

The responsivity to magnetic fields $\Bbar\supone$ is
\bea
\label{eq:S2SignalQSlope}
R(\omega) & \equiv & \frac{\partial  S_{2}\supQ(\omega)}{\partial \Bbar\supone(\omega)}  =
 \frac{ G S_1  \gamma F_+\supzero}
{-i \omega  + \Gamma + \Pbar}
\eea
such that
\bea
\label{eq:S2SignalQSlopeSq}
|R(\omega)|^2 & = & \frac{ ( G S_1  \gamma F_+\supzero )^2}
{\omega^2  + (\Gamma + \Pbar)^2}.
\eea

To find the quantum noise contribution to $S_{2}\supQ$, we use Eq.~(\ref{eq:FirstOrderDynamicsRes}) in Eq.~(\ref{eq:S2Signal2}) and assume that technical noise contributions to $S_{2+}$ and $F_+$ are negligible. Fluctuations of  $\Bbar\supone$, which represent a possible signal, are not counted as noise. We thus have the noise amplitude
\bea
\label{eq:S2SignalQNoise}
N_{S_2\supQ}(\omega) & = & {\cal I}[N_{S_2+}](\omega) +  \frac{G S_1  }
{-i\omega + \Gamma + \Pbar } {\cal I}[N_{F_+}](\omega).\hspace{9mm}
\eea

The two terms describe independent noise contributions, so that the noise power spectral density is 
\bea
\label{eq:S2SignalQNoise2}
{\cal S}_{S_2\supQ}(\omega) & = & {\cal S}_{{\cal I}[N_{S_2+}]}(\omega) +  \frac{G^2 S_1^2}{\omega^2 + (\Gamma + \Pbar)^2 } {\cal S}_{{\cal I}[N_{F_+}]}(\omega). 
\nonumber 
\eea 
Using the propagation of error formula, we get the magnetic power spectral density ${\cal S}_B(\omega)$, i.e. the square of the magnetic sensitivity.
\bea
\label{eq:Sensitivity2}
{\cal S}_B(\omega) & = &  |R(\omega)|^{-2}  |N_{S_2\supQ}(\omega)|^2
\nne \frac{ S_{{\cal I}[N_{F_+}]}(\omega)}{( \gamma F_+\supzero)^2}
+  \frac{\omega^2 + (\Gamma + \Pbar)^2 }{(G S_1\gamma F_+\supzero)^2} {\cal S}_{{\cal I}[N_{S_2+}]}(\omega). \hspace{8mm}
\eea

\subsection{Simplified notation}
To obtain the less cumbersome expressions used in the article text, we define the  in-phase signal amplitude
$\langle u \rangle \equiv G S_1 F_+\supzero$, the magnetic resonance line width $\Delta\omega \equiv \Gamma + \Pbar$, and the line-shape function 
\bea
{\cal L}(\omega) &\equiv& 
\frac{(\Gamma+\Pbar)^2}{\omega^2 + (\Gamma+\Pbar)^2}.
\eea
We can then express Eq.~(\ref{eq:S2SignalQSlopeSq}) as  

\bea
\label{eq:S2SignalQSlopeSqequiv}
|R(\omega)|^2 & = &  \gamma^2 \frac{\langle u \rangle^2}{\Delta \omega^2} {\cal L}(\omega).
\eea

The signal noise spectrum is ${\cal S}_v(\omega) \equiv {\cal S}_{S_2\supQ}(\omega)$, with an optical noise contribution
${\cal S}_{N_{S_2}} \equiv {\cal S}_{{\cal I}[N_{S_2+}]}(\omega)$. We note this is frequency independent, i.e. white noise.  The noise in ${\cal I}[N_{F_+}]$ is similarly white, and it is convenient to define its contribution to the signal as
\bea
\label{eq:SsigmaDef}
{\cal S}_\sigma \equiv \frac{G^2 S_1^2}{(\Gamma + \Pbar)^2} {\cal S}_{{\cal I}[N_{F_+}]} (\omega).
\eea 
The signal noise spectrum is then
\bea
\label{eq:SvInTermsOfSsigma}
{\cal S}_v(\omega) = {\cal S}_{N_{S_2}} + {\cal L} (\omega){\cal S}_\sigma.
\eea 
Using Eq.~(\ref{eq:S2SignalQSlopeSqequiv}) and Eq.~(\ref{eq:SvInTermsOfSsigma}) with the propagation of error formula we obtain the magnetic sensitivity in this notation:
\bea
\label{eq:Sensitivity2}
{\cal S}_B(\omega) & = &  |R(\omega)|^{-2} {\cal S}_v(\omega)
\nne \frac{\Delta \omega^2}{\gamma^2 \langle u \rangle^2} [ {\cal S}_\sigma + \frac{1}{{\cal L}(\omega)} {\cal S}_{N_{S_2}} ]
\nne \frac{1}{\gamma^2 \langle u \rangle^2} [\Delta \omega^2{\cal S}_\sigma +(\omega^2 +\Delta \omega^2){\cal S}_{N_{S_2}}],
\eea

c.f. Eq. (\ref{eq:Sensitivity}).

\section{Diffusion term}
\label{sec:FDT}

The fluctuation dissipation theorem relates the diffusion and relaxation terms in linear stochastic differential equations.  Often this is related to thermal noise, but here, since the spin system would relax to a maximum entropy state, i.e. a fully mixed state, the concept of temperature is not pertinent.  Nonetheless, as in the case of relaxation through loss of energy to a finite-temperature thermal reservoir, one can relate the diffusion to the equilibrium variance.

\newcommand{\cov}{{\rm cov}}

Meanwhile we can understand the spin noise by starting with Eq.~(\ref{eq:OriginalSpinDynamics}). Our question is: what does the diffusion term $\bN_{F}$ need to be, given the relaxation terms $-\Gamma \bF$ and $- P \bF$.  We note that the terms with $\bB$, $S_3$ and $P\hat{z} \Fmax$ are not relevant to this question - they will influence the dynamics over longer times, but the relation between $\bN_{F}$ and the relaxation terms must hold at every instant, and independently of the values of these other terms.  For that reason, it suffices to consider 
\bea
\label{eq:FDynOE}
\frac{d}{dt} F_i & = & -(\Gamma + P) F_i + \Sigma_{ij} \eta_j,
\eea
where $\eta_j$ are independent Gaussian white noise, defined by $\langle\eta_i(t) \eta_j(t')\rangle = \delta_{ij} \delta(t-t')$ and $\Sigma$ is a matrix.  This describes an Ornstein-Uhlenbeck process, i.e., Brownian motion with relaxation toward $\bF = {\bf 0}$. $\Sigma$ must satisfy the fluctuation-dissipation theorem, which is to say, give the correct equilibrium distribution for $\bF$. For a spin-$F$ system with small polarization, the equilibrium covariance matrix is $\cov(F_i, F_j) = \delta_{ij} N_A F(F+1)/3$ 
This is a diagonal matrix, indicating no correlations among different components $F_i$. It follows that each component $F$ is independently described by the same Ornstein-Uhlenbeck equation
\bea
\label{eq:FDynOE2}
\frac{d}{dt} F & = & - \Gamma' F + \sigma \eta,
\eea
where for convenience we have defined $\Gamma' \equiv \Gamma + P$. This has the well-known statistics $\langle F \rangle = 0$, $\langle F^2 \rangle = \sigma^2/2\Gamma'$ in the long-time limit.  This must equal $\NA{F(F+1)}/{3}$, and thus $\sigma^2 = 2 \Gamma' \NA{F(F+1)}/{3}$.  In this way we obtain  Eq.~(\ref{eq:NAutocorr}).

\bibliographystyleSM{./biblio/apsrev4-1no-url}
\bibliographySM{./biblio/PUMPINGBELLBLOOM}

\end{document}